\documentclass[prb,preprint,singlecolumn,amsmath,amssymb,superscriptaddress]{revtex4}
\usepackage[pdftex]{graphicx}
\usepackage{mathrsfs}
\usepackage[sort&compress]{natbib}

\newcommand{\be}{\begin{equation}}
\newcommand{\ee}{\end{equation}}
\newcommand{\bea}{\begin{eqnarray}}
\newcommand{\eea}{\end{eqnarray}}

\newcommand{\SI}{Supplementary Information}

\newcommand{\MIT}{Department of Physics, Massachusetts Institute of Technology, Cambridge, MA 02139, USA}
\newcommand{\NIMS}{Advanced Materials Laboratory, National Institute for Materials Science, 1-1 Namiki, Tsukuba 305-0044, Japan.}

\renewcommand{\phi}{\varphi}
\renewcommand{\epsilon}{\varepsilon}

\newcommand{\kp}{$k_\parallel$~}

\begin{document}
\title{Quantum and classical confinement of resonant states in a trilayer graphene Fabry-P\'{e}rot interferometer}
\author{L.~C. Campos}
  \affiliation{\MIT}
\author{A.~F. Young}
  \affiliation{\MIT}
\author{K. Surakitbovorn}
  \affiliation{\MIT}
\author{K. Watanabe}
  \affiliation{\NIMS}
\author{T. Taniguchi}
  \affiliation{\NIMS}

\author{P. Jarillo-Herrero}
  \affiliation{\MIT}
%\date{\today}
\maketitle

\textbf{The advent of few-layer graphenes\cite{Geim2007} has given rise to a new family of two-dimensional systems with emergent electronic properties governed by relativistic quantum mechanics. The multiple carbon sublattices endow the electronic wavefunctions with pseudospin, a lattice analog of the relativistic electron spin, while the multilayer structure leads to electric field effect tunable electronic bands. Here we use these properties to realize giant conductance oscillations in ballistic trilayer graphene Fabry-P\'{e}rot interferometers, which result from phase coherent transport through resonant bound states beneath an electrostatic barrier. We cloak these states\cite{Gu2011a} by selectively decoupling them from the leads, resulting in transport via non-resonant states and suppression of the giant oscillations. Cloaking is achieved both classically, by manipulating quasiparticle momenta with a magnetic field, and quantum mechanically, by locally varying the pseudospin character of the carrier wavefunctions. Our results illustrate the unique potential of trilayer graphene as a versatile platform for electron optics and pseudospintronics.}

The development of electronic devices with new functionality or improved performance depends on the ability to manipulate carrier degrees of freedom in low dimensional materials in new ways.  To this end, graphenes offer an appealing way forward: low electron-phonon coupling enables micron scale ballistic transport even at elevated temperatures \cite{Bolotin2008prl}, while the ambipolar tunability of the charge carrier density allows the realization of new device concepts based on electron optics\cite{Cheianov2007sc}. Moreover, the emergent pseudospin, which underlies many of the unique properties of graphitic electronic systems\cite{Ando1998,McEuen1999}, can be used as a basis for pseudospintronics\cite{Rycerz2007, Garcia-Pomar2008, San-Jose2009}, the lattice analog of spintronics\cite{Wolf2001}.  Among the family of few-layer graphenes,  ABA (or Bernal) stacked trilayer graphene (TLG) offers an ideal platform for exploring these effects, as the electronic structure consists of both monolayer and bilayer graphene (MLG and BLG)-like electronic bands\cite{Partoens2006, Latil2006, Koshino2009, CastroNeto2009, Taychatanapat2011,Bao2011}.  Crucially, the layered, two-dimensional arrangement of the carbon planes allows both the charge density and carrier ``flavor''---MLG- or BLG-like---to be varied through the application of out-of-plane electric fields\cite{Koshino2009}.  By varying the sum and difference of electric displacement fields generated by top and bottom electrostatic gates, the charge carriers in TLG can be tuned continuously between three regimes in which the Fermi level lies in a massless MLG-like hole band, a massive BLG-like ambipolar band, or a combination of the two (see \SI). In this Letter, we use this tunability to control the TLG band structure and pseudospin properties locally in an electronic Fabry-P\'{e}rot interferometer geometry in order to probe both inter- and intra-species scattering properties. Compared to previous work on pseudospin effects in quasi-ballistic graphene heterostructures \cite{Stander2009,Young2009}, which were limited to weak effects due to low sample quality, our devices exhibit a quantitatively and qualitatively different interference regime, both through a dramatic increase in sample quality and by exploiting the wide tunability of the TLG electronic structure.

Figures 1a-b show the device geometry, which consists of an ABA-stacked trilayer graphene flake encapsulated by hexagonal boron nitride (hBN).  The devices consist of three regions: two outer trilayer graphene leads (GLs), whose density is controlled by the global bottom gate, and the central locally gated region (LGR), where the density and perpendicular displacement field are independently controlled by the bottom and narrow top gate (see Supplementary Information). The LGR constitutes the electronic Fabry-P\'{e}rot cavity. Our hBN encapsulated samples show low disorder, with mobility $\mu\simeq$60,000 cm$^2$/V$\cdot$s at 300 mK. The resulting mean free path ($\ell_{\rm{mf}}\sim$600 nm) ensures that the quasiparticles are largely ballistic over length scales comparable to the top gate width ($\sim$60 nm).  Figure 1c shows the change in resistance of the LGR at zero magnetic field as a function of the voltages on the top ($V_{\rm{TG}}$) and bottom ($V_{\rm{BG}}$) gates.  The contact and GL resistance, which are independent of $V_{\rm{TG}}$, have been substracted from each $V_{\rm{BG}}$ trace for clarity.  The resulting resistance map is divided into four quadrants corresponding to different signs of the charge carrier density in the LGR and GLs \cite{Huard2007,Williams2007,Ozyilmaz2007prl}. The LGR resistance is higher in the bipolar regimes, II and IV, reflecting the presence of electrostatically created pn junctions, and negative in much of regions I and III, reflecting lowered resistance of the LGR when the absolute density is increased locally in the unipolar regime.  Figure 1d shows the derivative of the resistance with respect to top gate voltage, $\partial R/\partial V_{\rm{TG}}$.  Small resistance oscillations arising from phase coherent transport between the pn junctions are visible throughout the bipolar regime.  However, unlike electronic Fabry-P\'{e}rot resonances in MLG\cite{Young2011a}, the TLG data shows an additional set of resistance oscillations with much larger amplitude, visible even in the raw, undifferentiated data (white arrows in Fig. 1c).  In contrast to Fabry-P\'{e}rot oscillations in MLG, these oscillations are confined to a narrow region of the $V_{\rm{TG}}$$-V_{\rm{BG}}$ plane and are present only for very low magnetic fields (see Fig. 1e).  These giant oscillations were observed in multiple ABA TLG devices, and are a robust phenomenon, surviving to temperatures as high as $\sim50$~K (see \SI).

Finite element electrostatics simulations of the density and electric displacement field profile along the device, combined with a tight binding analysis of the ABA TLG band structure (see \SI), show that the giant oscillations are observed only when the Fermi level in the GLs lies in the isolated MLG-like hole band while that in the LGR lies in a BLG-like electron band (see Fig. 2a, inset).  This regime, which corresponds to an \textit{interflavor} MLG-BLG heterointerface, is not experimentally realizable in either mono- or bilayer graphene.  Below, we trace the origin of the giant oscillations to phase coherent transport through a small set of low transverse momentum resonant barrier states (RBS) in the LGR.  Remarkably, the presence of the giant oscillations in the conductance is sensitive to the pseudospin character of the charge carriers in the GLs.  Their decay with increasing GL density correlates with the population of a BLG-like band in the GLs.  In the resulting BLG-BLG interface, normally incident, forward propagating states on opposite sides of the pn junctions are pseudospin mismatched, and the RBS decouple from the GLs states and no longer manifest in transport.  In addition to this chirality-assisted cloaking of the RBS \cite{Gu2011a}, the RBS can also be cloaked classically using an external magnetic field, which induces a momentum mismatch between the RBS and the available states in the GLS.  We quantitatively describe both cloaking regimes, in which transport is diverted to other diffusive and ballistic channels and the giant oscillations disappear.

Oscillatory conductance in an electronic Fabry-P\'{e}rot etalon arises from quasiparticle trajectories that have comparable reflection and transmission amplitudes: too transmissive, and the particles are never trapped under the barrier, too reflective and they never enter from the lead.  At the same time, some selectivity in the angular transmission is necessary in order to avoid destructive phase averaging, which arises from the random injection angle of quasiparticles incident from the diffusive GLs.  Phase coherent transport across the LGR is thus strongly influenced by the transmission properties of the pn junctions defining the LGR, which determine the subset of available states through which ballistic transport can occur.  In a ballistic pn junction, the transmission amplitude, $T$, receives both a semiclassical and a fully quantum contribution, \cite{Cheianov2006,Allain2011,Young2011a}: $|T|\sim |T_{\rm{SC}}|\cdot| T_{\rm{Q}}|$. The semiclassical contribution, $|T_{\rm{SC}}|$, arises from the conservation of energy and the momentum parallel to the symmetric barrier, $k_\parallel$.  In a pn junction, the restricted Fermi surface at the center of the junction leads to purely evanescent transmission over the region where the local Fermi momentum\cite{Cheianov2006}, $|k_{\rm{F}}(x)|<|k_\parallel|$, resulting in preferential transmission of near normally incident quasiparticles with $|k_\parallel|\sim0$.

When electron wavefunctions do not have a pseudospin structure, $|T_{\rm{SC}}|$ generally suffices to describe barrier transmission.  In chiral electron systems, however, transmission also depends on the existence of a finite matrix element between incident and scattering states\cite{Katsnelson2006,Cheianov2006}.  Chiral particles have a fixed relationship between band index, momentum, and ``pseudospin''---the relative phase on the different carbon sublattices---leading to nontrivial contributions to $T_{\rm{Q}}$.  In MLG, for example, $T_{\rm{Q}}$ gives rise to the perfect transmission of quasiparticles normally incident on a pn junction (``Klein tunneling'') while in BLG the opposite occurs (``anti-Klein'' tunneling), with transmission forbidden at normal incidence\cite{Katsnelson2006,Tudorovskiy2012}.  Uniquely, TLG allows us to experimentally tune the chiral structure of the wavefunctions through the electric-field tunability of the band structure, allowing us to characterize the role of pseudospin in quasiparticle scattering.

Quantitative modeling of the giant oscillations, and their rapid disappearance as a function of both gate voltages and magnetic field, follows from the phase coherent Landauer formula\cite{Young2011a} for the conductance,
\begin{equation}
G_{\rm{pnp}}=\frac{4e^2}{h}\sum_{k_\parallel} \left|\frac{|T|^2}{1-\left(1-|T|^2\right)e^{i\rm{\theta}}}\right|^2
\label{eq_fp}
\end{equation}
where $\theta$ is the phase accumulated by carriers as they traverse the LGR\cite{Shytov2008prl,Young2011a}. We find that Eq. \ref{eq_fp} accounts very well for the conductance oscillation period and relative amplitude with respect to variation of the carrier density and magnetic field (see Figs. 3a-c, and discussion below). The only inputs needed are the electrostatic profile in the device (calculated by finite element analysis) and the TLG band structure\cite{Koshino2010,Taychatanapat2011} (see \SI), both obtained independently.

Analysis of Eq. 1 shows that two factors, absent in other graphene-based interferometers, conspire to enhance the amplitude of the giant oscillations.  First, the sum in (\ref{eq_fp}) runs over momenta corresponding to propagating states in both the GLs and LGR, $|k_\parallel|\leq\min(k_{\rm{F}}^{\rm{GL}},k_{\rm{F}}^{\rm{LGR}})$.  In the giant oscillation regime, the Fermi surface in the GLs is small in comparison to that in the LGR, leading to an effective collimation of the trajectories within the barrier.  Second, the interflavor nature of the MLG-BLG interface suppresses strong contributions to $T_{\rm{Q}}$, which is approximately angle independent, with transmissivity of $\sim 50\%$ (Fig. 2b).  The Fermi surface mismatch thus serves to preferentially inject carriers into the near-normal RBS in the junction, while the weak angular dependence and $\sim$1/2 value of the pseudospin matrix element ensures that transmission is in the optimal range, leading to the large observed oscillation amplitudes.

The giant oscillations are sensitive to small changes in the density in the GLs (see Fig 1c).  Figures 2a and 2c show fixed-$V_{\rm{BG}}$ conductance traces in and out of the giant oscillation regime.  Electrostatics simulations show that the sudden disappearance of the oscillations at high GL density coincides with a transition in the pseudospin structure of the GL wavefunctions: the oscillations are quickly suppressed once the BLG-like bands start to become occupied in the GLs (see insets in Fig. 2a and 2c).  The nature of this suppression can be traced to the effect of BLG anti-Klein tunneling\cite{Katsnelson2006,Gu2011a, Allain2011} on $T_{\rm{Q}}$.  Whereas for the MLG-BLG interface, carriers are injected predominantly into the low $k_{\parallel}$  RBS states, the BLG-BLG interface features a chiral mismatch between low $k_{\parallel}$ states on opposite sides of the barrier.  Figures 2b-d show numerical simulations of the pn junction transmission amplitudes for values of the GL density corresponding to the traces in Fig. 2a-c.  The qualitative change in the total pn junction transmission is visible as a shift from high transmission to near-complete reflection at normal incidence: in the BLG-BLG regime, the RBS are effectively cloaked by chiral mismatch to the leads.  While it is the tunable carrier chirality in ABA TLG that makes this effect observable in the current experiment, ``anti-Klein'' tunneling is a generic feature of chiral carriers with 2$\pi$ Berry phase such as those in BLG.

Cloaking of the RBS can also be achieved by a classical mechanism through the application of an out of plane magnetic field, $B$.  Figure 3b shows the $B$ dependence of the giant oscillations: they disappear rapidly with $B$, and are completely suppressed by $B\sim$100 mT. This is in stark contrast to the Fabry-P\'{e}rot oscillations observed in MLG \cite{Young2009,Young2011a}, and the smaller Fabry-P\'{e}rot oscillations faintly visible in Figs. 1d-e, which survive up to high $B$ (see \SI). The disappearance of the giant oscillations with $B$ can be tied to a classical version of the cloaking of the RBS, facilitated in TLG by the widely tunable Fermi surface area.  The magnetic field exerts a Lorentz force on particles traversing the LGR (see Fig. 4).  During the resulting cyclotron motion, \kp is no longer conserved as the carriers cross the LGR, and changes with coordinate\cite{Gu2011}: $k_\parallel(x_2)=k_\parallel(x_1)-eBL/\hbar$, where $x_1$ and $x_2$ denote the positions of the two pn junctions and $L=x_2-x_1$.  However, with the exception of small corrections \cite{Shytov2009ssc}, $k_\parallel$ is still conserved in the tunneling process at a single pn junction.  Classical cloaking occurs when this condition can no longer be satisfied (see Fig. 4c).  Because the giant oscillations are observed in a regime of extreme Fermi surface mismatch ($|k_{\rm{F}}^{\rm{LGR}}|\gg|k_{\rm{F}}^{\rm{GL}}|$), only a small magnetic field is necessary to make simultaneous momentum and energy momentum conservation impossible at the second interface, $k_\parallel(x_2)>k_{\rm{F}}^{\rm{GL}}$.  Figure 4 shows the three distinct magnetic field regimes schematically. For small values of $B$, the field simply introduces a difference in the incident angle of a quasiparticle trajectory on the two junctions for most values of $k_\parallel$(see Fig. 4b).  However, above a critical magnetic field, $B_{\rm{C}}\sim2\hbar k_{\rm{F}}^{\rm{GL}}/(eL)$, the field imparts such a large transverse momentum to the carriers in the RBS that they can no longer escape to the GLs, and effectively decouple from transport (see Fig. 4c).  As in the case of chirality assisted cloaking described above, the resulting decoupling of the RBS from the lead states drives transport to diffusive channels, in which impurity scattering in the LGR provides the momentum relaxation required for quasiparticle escape.

ABA TLG constitutes a high mobility and highly tunable electronic system: carrier density, Fermi surface area, energy-momentum dispersion, quasiparticle chirality and pseudospin can all be tuned electrostatically over large ranges. In this Letter, we have used this tunability to realize a novel Fabry-P\'{e}rot interferometer device, where we explore both quantum confinement of chiral carriers as well as the limits of semiclassical magnetoconfinement in nanostructures.  These phenomena and the unique versatility of TLG hold promise for the realization of devices with novel functionality based  on pseudospintronics.

\section{Acknowledgements}
We acknowledge discussions with Nan Gu, Leonid Levitov, Thiti Taychatanapat, Valla Fatemi and Javier Sanchez-Yamagishi, as well as Philip Kim, Leonid Levitov and Amir Yacoby for comments on the manuscript. This work was financially supported by the Office of Naval Research GATE MURI and National Science Foundation Career Award DMR-0845287. L.C.C. acknowledges partial support by the Brazilian agency CNPq. This research has made use of the NSF funded MIT CMSE and Harvard CNS facilities.

\section{Contributions}
L.C.C. and K.S. fabricated the devices.  K.W. and T.T. synthesized the hBN crystals. L.C.C. performed the measurements.  L.C.C., A.F.Y. and P.J-H. analyzed the data and co-wrote the paper.

\section{Additional information}
Correspondence and requests for materials should be addressed to P.J-H. (email: pjarillo@mit.edu).

%\bibliography{references}
%\bibliographystyle{naturemag1}

\begin{figure*}[ht]
\includegraphics[width=170mm]{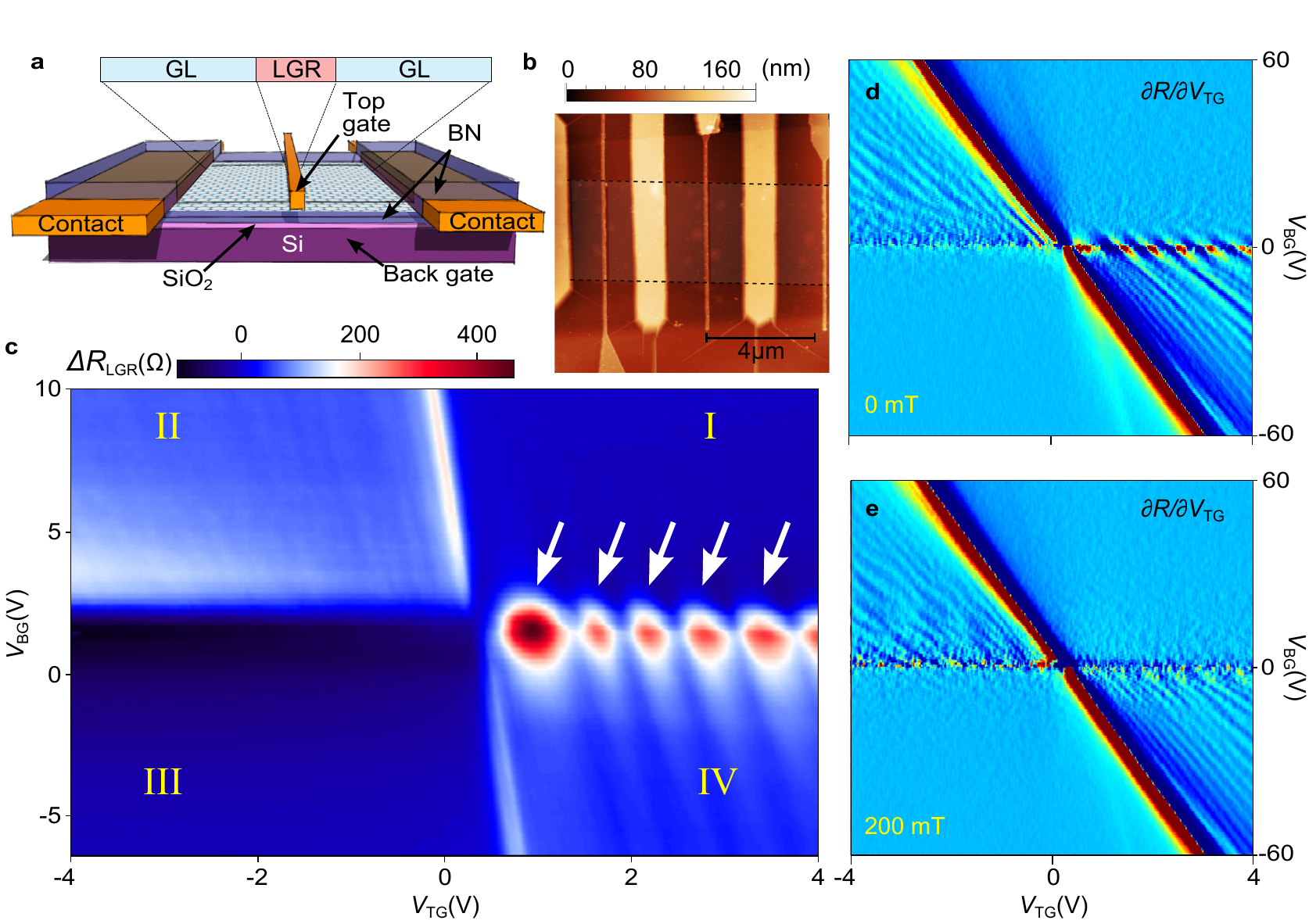}
\caption{\textbf{Trilayer graphene heterojunction device schematic and electronic transport measurements.}
\textbf{a}, Schematic device representation.  A narrow top gate is fabricated on an hexagonal boron nitride encapsulated ABA trilayer graphene flake.  A global highly doped Si bottom gate controls electron density and the electrical displacement throughout the entire flake, while the top gate affects only the locally gated region (LGR).  Interfaces between regions of different carrier type can be induced electrostatically at the LGR boundaries by appropriate choice of the top and bottom gate voltages.
\textbf{b}, Atomic force micrograph of the measured device. The black dashed lines indicate the TLG edges.
\textbf{c}, Two terminal resistance data acquired at \emph{T}=300 mK and \emph{B}=0T.  The resistance of the leads has been subtracted by removing a constant resistance corresponding to a uniform channel: $\Delta R_{\rm{LGR}}(V_{\rm{TG}},V_{\rm{BG}})=\emph{R}(V_{\rm{BG}},\emph{V}_{\rm{TG}})-\emph{R}(V_{\rm{BG}},0.37 \rm{V})$.  Large amplitude oscillations are visible when the LGR is negatively doped (arrows).  The oscillations decay rapidly for large absolute density in the GLs.
\textbf{d}, Numerical derivative of the two terminal resistance at \emph{B}=0. In addition to the giant oscillations, additional resonances with smaller amplitude are visible throughout the bipolar regions, II and IV.
\textbf{e}, Numerical derivative of the two terminal resistance at \emph{B}=200 mT.  While the small oscillations persist, the giant oscillations are completely suppressed by the classical cloaking effect discussed in the main text.}
  \label{f1}
\end{figure*}

\thispagestyle{empty}
\begin{figure*}[ht]
  \includegraphics[width=150mm]{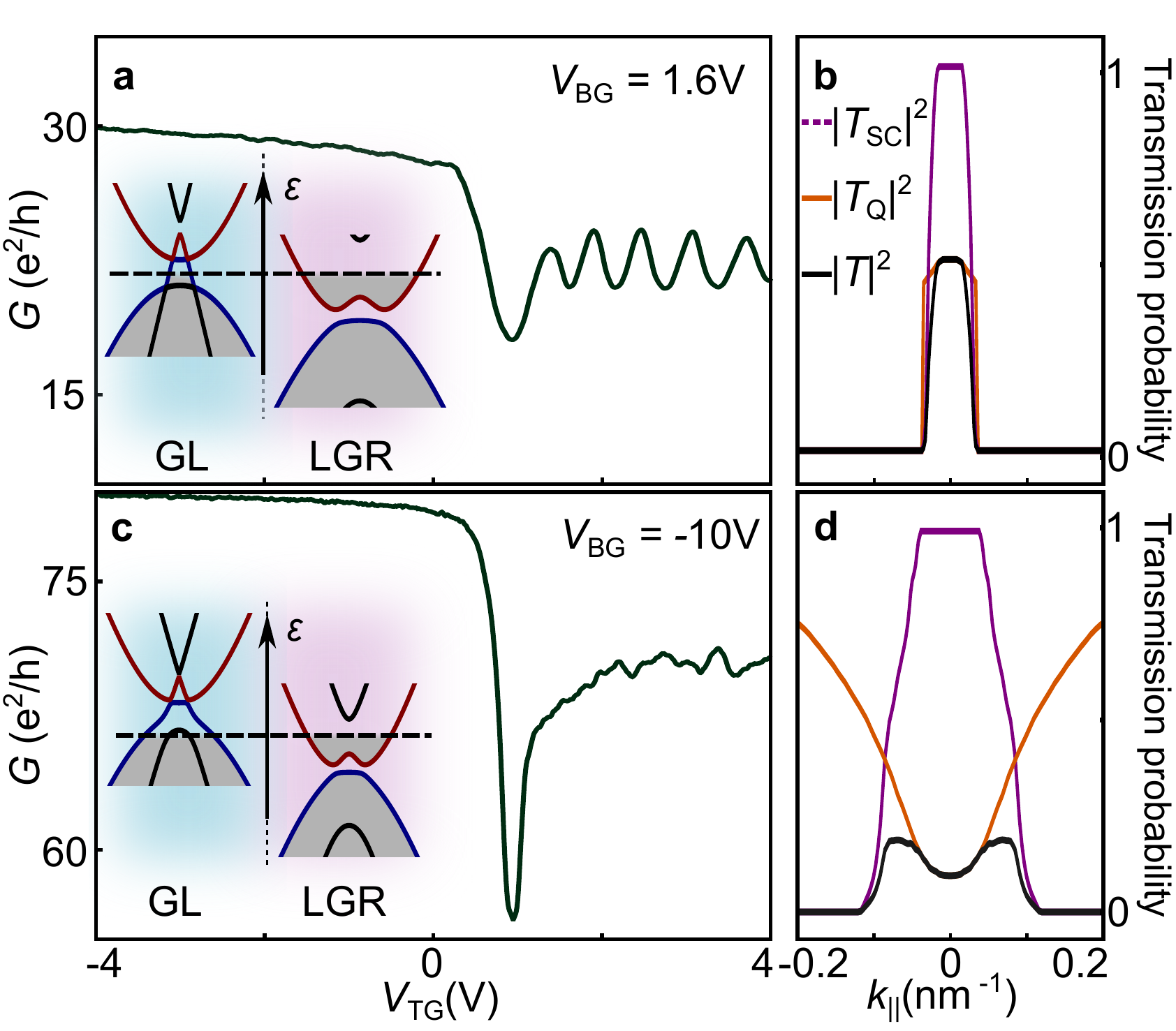}
  \caption{\textbf{Quantum cloaking of the resonant barrier states}.
\textbf{a}, Conductance, $G$, as function of of \emph{V}$_{\rm{TG}}$ at \emph{V}$_{\rm{BG}}$=-1.6 V, showing large amplitude oscillations.  The oscillations are observed when the GLs are populated by MLG-like p-type carriers, while the LGR contains BLG-like n-type carriers (inset to \textbf{a}).
\textbf{b}, Calculated transmission probability at a single pn interface in the giant oscillation regime.  Interspecies quantum matrix elements do not have a strong $k_\parallel$ dependence, and the Fermi surface mismatch injects carriers preferentially into the low $k_\parallel$ RBS, leading to the giant oscillations.
\textbf{c}, The chirality-assisted cloaking regime, in which the interface separates BLG-like electron and hole doped regions (inset).  The measured conductance shows only small amplitude fluctuations (see also \SI).
\textbf{d}, Calculated transmission probability at a single pn interface for the electrostatic conditions in \textbf{c}.  In this regime, $|T|$ is dominated by BLG-BLG anti-Klein tunneling: the pseudospin mismatch between $k_\parallel= 0$ holes in the GL and $k_\parallel= 0$ electrons in the LGR suppresses transmission through the RBS\cite{Gu2011}, resulting in a dip in the normal transmission probability and ultimately the suppression of the giant oscillations.
  }\label{f2}
\end{figure*}

\newpage
\thispagestyle{empty}
\begin{figure*}[ht]
  \includegraphics[width=170mm]{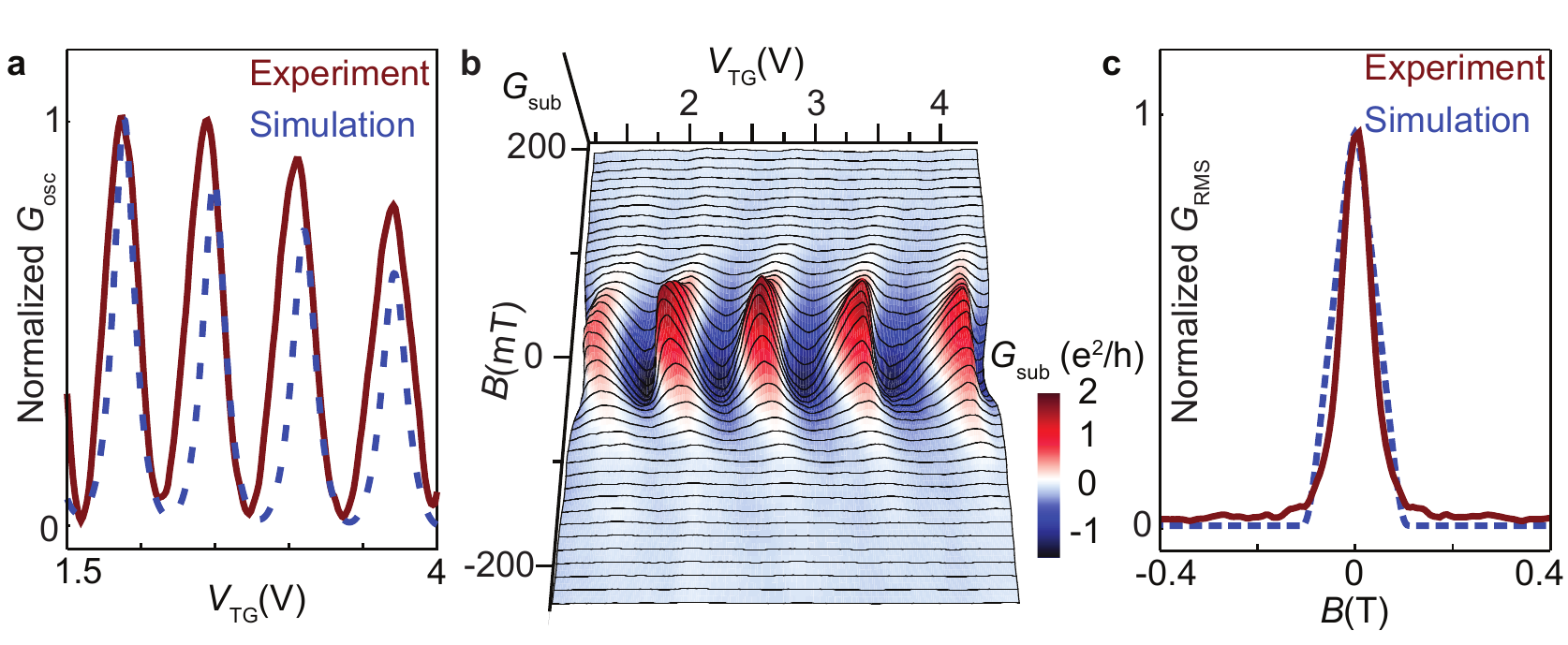}
  \caption{\textbf{Classical cloaking of the RBS}
  \textbf{a},  Comparison between measured (solid) and simulated (dashed) normalized conductance oscillations in the RBS channel at zero magnetic field (see \SI).
  \textbf{b}, Magnetic field dependence of the giant oscillations at \emph{V}$_{\rm{BG}}$=1.6 V.  The oscillations decay rapidly with magnetic field, indicating a shift to transport channels that do not include the RBS.  The average conductance over the displayed gate voltage range has been subtracted from each constant \emph{B} trace to remove the strong magnetic field dependence of the GL resistance.  \textbf{c} Comparison of the normalized amplitude of the measured and simulated conductance oscillations as a function of magnetic field. The simulations reproduce the observed rapid decay in oscillation amplitude due to classical confinement, in which collapse of the giant oscillations occurs as $B\rightarrow B_{\rm{C}}=2\hbar k_{\rm{F}}^{\rm{LGR}}/(eL)$.
  }
  \label{f3}
\end{figure*}

\newpage
\thispagestyle{empty}

\begin{figure*}[ht]
  \includegraphics[width=150mm]{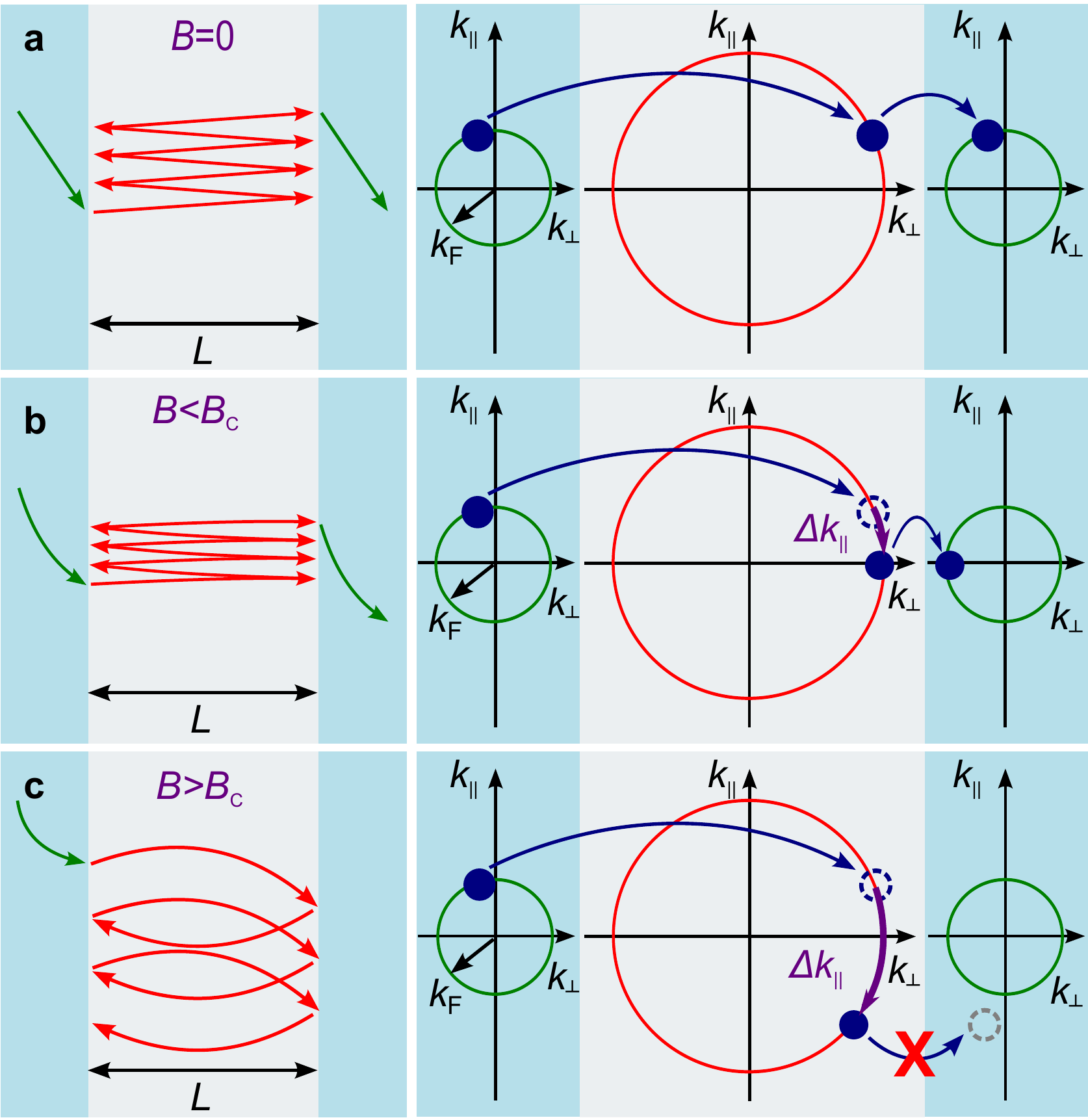}
  \caption{\textbf{Schematic illustration of the classical magneto-cloaking of the RBS.}
  \textbf{a}, At \emph{B}=0, carriers are injected from the GL Fermi surface into low \kp states in the LGR, resulting in transport through the RBS and giant oscillations in the resistance.
  \textbf{b}, As the magnetic field increases, quasiparticles gain $\Delta k_{\rm{\parallel}}=eBL/\hbar$ as they cross the LGR. While trajectories for which $\Delta k_{\parallel}>k_{\rm{F}}^{\rm{GL}}$  are excluded from ballistic transport, some ballistic transport is still possible via the RBS.
  \textbf{c},   For $B>B_{\rm{C}}=2\hbar k_{\rm{F}}^{\rm{GL}}/eL$, ballistic transport via the RBS is no longer possible.  Charge transport proceeds by momentum nonconserving channels that do not involve the RBS, likely mediated by impurity scattering within the LGR, and the giant oscillations disappear.}\label{f4}
\end{figure*}

\end{document}